\documentclass[sigconf]{acmart}
\usepackage{tikz}
\usetikzlibrary{positioning, arrows.meta, fit, backgrounds, calc, shapes.geometric}
\setcopyright{none}
\renewcommand\footnotetextcopyrightpermission[1]{}
\pdfoutput=1
\setcopyright{none}
\renewcommand\footnotetextcopyrightpermission[1]{}
\acmConference[AgenticSE @ CAIS '26]{Workshop on Agentic Software Engineering}{May 26, 2026}{San Jose, CA, USA}
\acmYear{2026}
\acmBooktitle{Workshop on Agentic Software Engineering (AgenticSE), co-located with ACM CAIS 2026}

\begin{document}

\title{Monitoring Agentic Systems Before They're Reliable}\thanks{Accepted to the Workshop on Agentic Software Engineering (AgenticSE), co-located with ACM CAIS 2026.}

\author{Marisa Ferrara Boston}
\authornote{Corresponding author.}
\email{marisa@reinsai.com}
\orcid{0009-0003-0943-2491}
\affiliation{%
  \institution{Reins AI}
  \country{USA}
}

\author{Glen Hanson}
\email{glen@reinsai.com}
\orcid{0009-0001-1637-3616}
\affiliation{%
  \institution{Reins AI}
  \country{USA}
}

\author{Effi Georgala}
\email{effi@reinsai.com}
\orcid{0009-0001-2724-7439}
\affiliation{%
  \institution{Reins AI}
  \country{USA}
}

\author{JD Hudgens}
\email{jd@reinsai.com}
\affiliation{%
  \institution{Reins AI}
  \country{USA}
}

\author{Heather Frase}
\email{hnfrase@veraitechus.com}
\orcid{0009-0004-9161-0716}
\affiliation{%
  \institution{Veraitech}
  \country{USA}
}

\begin{abstract}
Agentic systems entering production typically operate as partially integrated assemblies where structural defects, not task-level errors, dominate the failure landscape. At this maturity level, task-level error detection may be infeasible: structural failure modes mask the signal that task-level monitors are designed to detect.

We present a monitoring and triage methodology that decomposes agentic system evaluation into three dimensions (quality, suitability, efficiency) at three monitoring scopes (within-run, cross-run, structural), using variance as a characterization signal. Findings are routed through severity classification adapted from FMEA, concentrating human attention on the subset that warrants investigation. We evaluate the methodology on a synthetic testbed of 220 runs across 120 document bundles with controlled error injection, processed by an early-stage system with known integration defects.

Three results emerge. First, monitor scope determines failure type: within-run monitors surface deterministic stage defects (CV = 0.02), cross-run monitors surface stochastic integration consequences (CV = 1.25, 24\% at severity L2), and a structural monitor identifies an integration gap with perfect consistency (CV = 0.00). Second, injected task-level errors are indistinguishable from clean baselines, confirming that structural defects mask task-level signal. Third, deterministic triage routes 97\% of findings to automated tracking, concentrating human investigation on the 2\% reflecting variable system behavior.

For this system, effective monitoring begins with structural diagnosis before error detection becomes viable. We propose, based on Stage~1 evidence, a maturity-staging model in which monitoring transitions from structural characterization to error detection to reliability tracking as integration defects are resolved. The taxonomy, CV-based scope characterization, and severity model transfer architecturally to document-driven, multi-stage agentic workflows in regulated industries; specific calibrations are domain-specific. Deploy monitoring early: the first thing it finds is the most important thing to fix.
\end{abstract}

\maketitle

\section{Introduction}
\label{sec:introduction}

Agentic AI systems are entering production across regulated industries
(audit, finance, healthcare, legal services) where failures carry
strategic, financial, and reputational consequences. Before
deployment, these systems are typically validated through capability
benchmarks~\cite{jimenez2024swebench}: accuracy on held-out test
sets, compliance with policy constraints, or adversarial robustness
under red-team evaluation. Yet once deployed, the question shifts
from ``does the model work?'' to ``when does the system remain
reliable, and when doesn't it?''

Current monitoring practice has not kept pace with this shift. The
dominant approach tracks aggregate performance metrics (mean
accuracy, composite quality scores, error rates averaged over time)
and alerts when a threshold is crossed. This works well for narrow,
well-scoped tasks with reliable ground truth and homogeneous inputs.
Agentic systems in regulated domains violate all three assumptions.
Tasks are heterogeneous: an audit agent may process a
straightforward bank reconciliation in one session and a complex
multi-document cutoff analysis in the next. Scoring is uncertain:
LLM-based evaluators exhibit variance across runs, and human
reviewers disagree on edge cases. Ground truth is partial or absent:
for many agent outputs, correctness is a matter of professional
judgment rather than binary verification.

Under these conditions, mean-based monitoring creates false
confidence. A system reporting 87\% mean accuracy may be performing
reliably on routine tasks while failing unpredictably on the complex
cases that carry the highest operational risk. The problem is not
merely imprecision. It is a systematic bias: aggregate metrics
disproportionately reflect performance on the frequent, easy cases
that dominate the distribution, while obscuring the rare,
high-severity failures where audit and business risk actually
concentrate. Recent work supports this concern. Rabanser et
al.~\cite{rabanser2026reliability} demonstrate that compressing agent
behavior into a single success metric obscures critical operational
properties across consistency, robustness, predictability, and safety
dimensions, and that recent capability gains have yielded only small
reliability improvements. Our work addresses the operational side of
this gap: given that aggregate metrics are insufficient, how should
production monitoring systems detect and prioritize the failures that
matter most?

Even when monitoring surfaces anomalies, the process typically stops at detection. Without severity-calibrated triage, analysts review low-severity anomalies at the same cost as critical failures, and monitoring generates evidence that is necessary but insufficient for reliable operation.

Most agentic systems enter production as partially integrated assemblies where structural integration defects, not task-level errors, dominate the failure landscape. Monitoring these systems requires a methodology that produces value at this early stage, surfacing the integration defects that must be resolved before task-level error detection becomes viable.

This paper addresses these gaps. We present a monitoring and
automated triage methodology that decomposes agentic system behavior
across three evaluative dimensions (quality, suitability, efficiency)
at three monitoring scopes (within-run, cross-run, structural), with
variance as a signal for characterizing system behavior. Findings are
routed through FMEA-based severity classification calibrated to
operational impact, concentrating human attention on the subset of
findings that warrant investigation.

\smallskip
\noindent Our contributions are:

\begin{enumerate}
    \item A \textbf{triangulated monitoring methodology} that
    decomposes agent evaluation into three complementary dimensions
    (quality, suitability, and efficiency) at three monitoring scopes
    (within-run, cross-run, structural), with within-category variance
    as a first-class signal for characterizing system behavior
    alongside aggregate means.

    \item An \textbf{FMEA-based automated triage system} that
    classifies detected failures by operational severity using
    structured metadata from triangulated monitoring, adapting Failure
    Mode and Effects Analysis methodology from safety-critical
    industries to agentic AI systems.

    \item \textbf{Empirical evidence that monitor scope determines
    failure type and severity profile.} On a synthetic testbed with
    known ground truth (120 document bundles, 220 runs, five error
    subtypes at four difficulty levels), within-run monitors produce
    high-volume, uniformly low-severity (L3) output reflecting
    deterministic stage defects. Cross-run monitors produce sparse
    output with 24\% of findings at severity L2 (Critical), surfacing
    stochastic consequences of integration gaps. A structural monitor
    identifies an integration integrity defect with complete
    consistency across all runs. Variance across monitor outputs
    (CV range: 0.00--1.25) characterizes these scope classes without
    ground truth labels.

    \item \textbf{Evidence that scope-aware triage achieves
    substantial review reduction.} Deterministic triage rules route
    all 10{,}210 L3 findings to automated monitoring and all 243 L2
    findings to human investigation, a 43$\times$ reduction in
    analyst volume. The contribution is the monitoring architecture
    and severity design that make deterministic routing sufficient,
    eliminating per-incident human classification.
\end{enumerate}

\noindent We validate the methodology on a synthetic testbed designed
for the Subsequent Events and Unrecorded Liabilities (SURL) audit
procedure under US GAAP, representative of the broad class of
document-driven agentic workflows across regulated industries. The
testbed uses realistic document bundles with controlled error
injection, processed by an early-stage agentic system with known
integration defects. Observed patterns are consistent with integration-stage behaviors observed qualitatively in practitioner deployments of agentic systems in regulated industries, though no production data is reported here.

\subsection{Related Work}
 
Capability benchmarks such as SWE-bench~\cite{jimenez2024swebench} and
AgentBench~\cite{liu2024agentbench} evaluate whether language models can
act as agents in interactive environments. These benchmarks address model
selection---can this LLM reason over multi-step tasks?---but do not address
the question our work targets: whether a deployed agentic \emph{system}
remains reliable in production, where the failure landscape is dominated by
integration defects rather than model limitations.
 
The agent observability ecosystem has matured rapidly. OpenTelemetry's
GenAI semantic conventions now define standardized spans for model
invocations and agent operations~\cite{otel2025genai}, with active
proposals extending coverage to agentic tasks, tool calls, and
remediation workflows~\cite{otel2025agents}. Commercial platforms
(Phoenix/Arize, Langfuse, LangSmith, Maxim, Azure AI Foundry) provide
tracing and general-purpose evaluation infrastructure built on these
standards: latency, token consumption, trace completion, drift
detection, and configurable LLM-based judges. These platforms are
necessary components of production AI operations and serve as
integration targets for the methodology presented here.

Our work differs from these platforms along three axes. First, on
\textit{detection signal}: standard practice tracks aggregate
performance against threshold alerts, while we use within-category
variance as a first-class signal that distinguishes monitor scope
classes without requiring ground truth labels
($\S$\ref{sec:variance-method}). Second, on \textit{scope
decomposition}: existing platforms treat findings as a flat stream,
while we partition monitors into within-run, cross-run, and
structural scopes that surface categorically different failure
classes. Third, on \textit{triage}: existing platforms surface
findings without severity calibration, leaving prioritization to
human reviewers; we route findings through deterministic
FMEA-based severity classification, eliminating per-incident
classification overhead. The methodology is complementary, not
competitive: findings flow back into platform telemetry as
structured span metadata. A recent Brookings--CMU--Berkeley
initiative on agentic AI evaluation identifies this
measurement-to-governance gap as a research priority~\cite{brookings2026agentic}.
 
Rabanser et al.~\cite{rabanser2026reliability} demonstrate that compressing
agent behavior into single success metrics obscures critical operational
properties across consistency, robustness, and safety dimensions, motivating
our multi-dimensional evaluation design. The closest methodological
precedent is the application of FMEA to AI/ML systems. Recent work applies
Process FMEA to ML training pipelines~\cite{schmitt2025mlfmea} or uses LLMs
to automate traditional FMEA workflows~\cite{younus2025fmea}. Our
contribution differs in target and timing: we apply FMEA to the
\emph{operational monitoring output} of deployed agentic systems, using it
as a runtime triage mechanism rather than a design-time risk identification
tool.

The remainder of this paper is organized as follows.
Section~\ref{sec:methodology} presents the triangulated evaluation
methodology and the role of variance as a detection and
characterization signal. Section~\ref{sec:triage} describes the
FMEA-based automated triage system. Section~\ref{sec:setup} details
the experimental setup, including the synthetic testbed and agent
under test. Section~\ref{sec:results} reports results on monitor
firing profiles, the null result on error detection, and triage
routing. Section~\ref{sec:discussion} discusses maturity staging,
generalizability, limitations, and the connection to continuous
reliability improvement.
\raggedbottom
\section{Triangulated Evaluation Methodology}
\label{sec:methodology}

As established in Section~1, standard aggregate monitoring assumes
task homogeneity, scoring reliability, and clean ground truth, all
of which break down for agentic systems in regulated domains. This
section presents our response: a monitoring methodology that
decomposes evaluation into three complementary dimensions and uses
within-category variance as a first-class signal for characterizing
system behavior.
\raggedbottom
\subsection{Three Evaluative Dimensions}

We decompose agent evaluation into three dimensions, each capturing
a distinct class of failure:

\textbf{Quality} measures whether the agent produced a correct
result. Evaluators in this dimension assess output accuracy,
completeness, information grounding, and the absence of hallucinated
content. An agent that extracts the wrong invoice amount or
fabricates a document reference fails on quality regardless of how it
arrived at the answer.

\textbf{Suitability} measures whether the agent performed the task
appropriately. Evaluators assess instruction adherence, contextual
fit, interaction patterns, and alignment with the intended workflow.
An agent that produces an accurate answer but ignores the auditor's
specific instructions, bypasses required verification steps, or
provides guidance misaligned with the engagement context fails on
suitability even when its output is technically correct.

\textbf{Efficiency} measures whether the agent delivered value
relative to cost. Evaluators assess task completion time,
computational cost (e.g., token usage), and the human review burden
imposed by the system's output. An agent that produces high-quality,
suitable results but requires extensive human correction or consumes
disproportionate resources may not justify deployment.

These three dimensions are complementary, not redundant. An agent
can be accurate but unsuitable (correct answer, wrong process),
suitable but inefficient (right process, excessive cost), or
efficient but low-quality (fast but wrong). Monitoring each dimension
independently surfaces failure modes that a single composite score
would mask.

\begin{figure*}[t]
\centering
\resizebox{\textwidth}{!}{%
\begin{tikzpicture}[
    node distance=0.6cm and 0.8cm,
    font=\sffamily\small,
    >={Stealth[length=3pt]},
    source/.style={
        rectangle, rounded corners=3pt, draw=gray!60, fill=gray!8,
        minimum height=0.7cm, minimum width=2cm, align=center,
        font=\sffamily\footnotesize
    },
    evaluator/.style={
        rectangle, rounded corners=3pt, draw=blue!50!black!60, fill=blue!6,
        minimum height=0.7cm, minimum width=1.6cm, align=center,
        font=\sffamily\footnotesize
    },
    monitor/.style={
        rectangle, rounded corners=3pt, draw=teal!70!black, fill=teal!8,
        minimum height=0.7cm, minimum width=1.8cm, align=center,
        font=\sffamily\footnotesize
    },
    triage/.style={
        rectangle, rounded corners=3pt, draw=orange!70!black, fill=orange!6,
        minimum height=0.7cm, minimum width=1.8cm, align=center,
        font=\sffamily\footnotesize
    },
    output/.style={
        rectangle, rounded corners=3pt, draw=red!60!black, fill=red!5,
        minimum height=0.7cm, minimum width=1.8cm, align=center,
        font=\sffamily\footnotesize
    },
    groupbox/.style={
        rectangle, rounded corners=6pt, draw=#1!40, fill=#1!3,
        inner sep=8pt
    },
    arrow/.style={->, thick, gray!70},
    bigarrow/.style={->, very thick, black!50, line width=1.5pt},
    label/.style={font=\sffamily\scriptsize\bfseries, text=black!60},
]


\node[source] (traces)       {Traces};
\node[source, below=0.4cm of traces] (interactions) {Interactions};
\node[source, below=0.4cm of interactions] (outputs)  {Outputs};

\begin{scope}[on background layer]
    \node[groupbox=gray, fit=(traces)(interactions)(outputs),
          label={[label, anchor=south]above:Agent Telemetry}] (srcgroup) {};
\end{scope}

\node[evaluator, right=1.5cm of traces] (rules) {Rules};
\node[evaluator, below=0.4cm of rules] (stats) {Statistical};
\node[evaluator, below=0.4cm of stats] (llm)   {LLM Judges};

\begin{scope}[on background layer]
    \node[groupbox=blue!50!black, fit=(rules)(stats)(llm),
          label={[label, anchor=south]above:Evaluators}] (evalgroup) {};
\end{scope}

\node[monitor, right=1.5cm of rules] (quality)    {Quality};
\node[monitor, below=0.4cm of quality] (suitability) {Suitability};
\node[monitor, below=0.4cm of suitability] (efficiency) {Efficiency};

\node[monitor, below=0.5cm of efficiency, draw=teal!90!black, fill=teal!15,
      minimum width=1.8cm, font=\sffamily\footnotesize\itshape] (variance) {+ Variance signals};

\begin{scope}[on background layer]
    \node[groupbox=teal!70!black, fit=(quality)(suitability)(efficiency)(variance),
          label={[label, anchor=south]above:Monitors}] (mongroup) {};
\end{scope}

\foreach \s in {traces, interactions, outputs} {
    \foreach \e in {rules, stats, llm} {
        \draw[arrow, gray!30, thin] (\s.east) -- (\e.west);
    }
}

\draw[arrow, gray!30, thin] (rules.east) -- (quality.west);
\draw[arrow, gray!30, thin] (stats.east) -- (suitability.west);
\draw[arrow, gray!30, thin] (llm.east) -- (efficiency.west);
\draw[arrow, gray!30, thin] (rules.east) -- (suitability.west);
\draw[arrow, gray!30, thin] (stats.east) -- (quality.west);
\draw[arrow, gray!30, thin] (llm.east) -- (quality.west);
\draw[arrow, gray!30, thin] (llm.east) -- (suitability.west);

\coordinate (bridge_start) at ($(mongroup.east)+(0.3,0)$);
\coordinate (bridge_end) at ($(mongroup.east)+(1.2,0)$);
\draw[bigarrow] (bridge_start) -- (bridge_end);


\node[triage, right=2.0cm of quality] (classify) {Failure\\Classification};
\node[triage, right=0.8cm of classify] (risk) {Risk\\Classification};
\node[triage, right=0.8cm of risk] (prioritize) {Failure\\Prioritization};
\node[output, right=0.8cm of prioritize] (response) {Investigation\\$\&$ Response};

\node[below=0.35cm of risk, font=\sffamily\tiny, text=black!50, align=center] (sevlabels)
    {L1-Cata.~|~L2-Crit.~|~L3-Marg.~|~L4-Negl.};

\node[below=0.05cm of sevlabels, font=\sffamily\tiny\itshape, text=black!40] (riskformula)
    {Severity $\times$ Probability $\times$ Exposure};

\node[below=0.35cm of response, font=\sffamily\tiny, text=black!50, align=center] (resptypes)
    {Corrective Action\\Workaround~|~Patch};

\begin{scope}[on background layer]
    \node[groupbox=orange!70!black,
          fit=(classify)(risk)(prioritize)(response)(sevlabels)(riskformula)(resptypes),
          label={[label, anchor=south]above:FMEA-Based Triage}] (triagegroup) {};
\end{scope}

\draw[arrow, orange!60!black] (classify.east) -- (risk.west);
\draw[arrow, orange!60!black] (risk.east) -- (prioritize.west);
\draw[arrow, orange!60!black] (prioritize.east) -- (response.west);

\node[above=0.6cm of prioritize, font=\sffamily\tiny, text=black!50,
      rectangle, rounded corners=2pt, draw=black!20, fill=yellow!8,
      inner sep=3pt] (human) {Human confirms};
\draw[->, thin, black!30, dashed] (human.south) -- (prioritize.north);

\node[below=0.6cm of srcgroup.south, font=\sffamily\small\bfseries, text=black!50]
    {(a)~Triangulated Monitoring};
\node[below=0.6cm of triagegroup.south, font=\sffamily\small\bfseries, text=black!50]
    {(b)~Automated Triage};

\end{tikzpicture}%
}

\caption{Combined monitoring and triage architecture. \textbf{(a)}~Agent telemetry feeds rule-based, statistical, and LLM-based evaluators that produce scores across three dimensions (quality, suitability, efficiency), with variance signals as a primary detection mechanism. \textbf{(b)}~Structured metadata from monitoring feeds FMEA-based triage: failures are classified, assessed for risk (severity $\times$ probability $\times$ exposure, calibrated to MIL-STD-882E categories), prioritized, and routed to appropriate response types. Human review is triggered for L1/L2 severity.}
\label{fig:architecture}
\end{figure*}
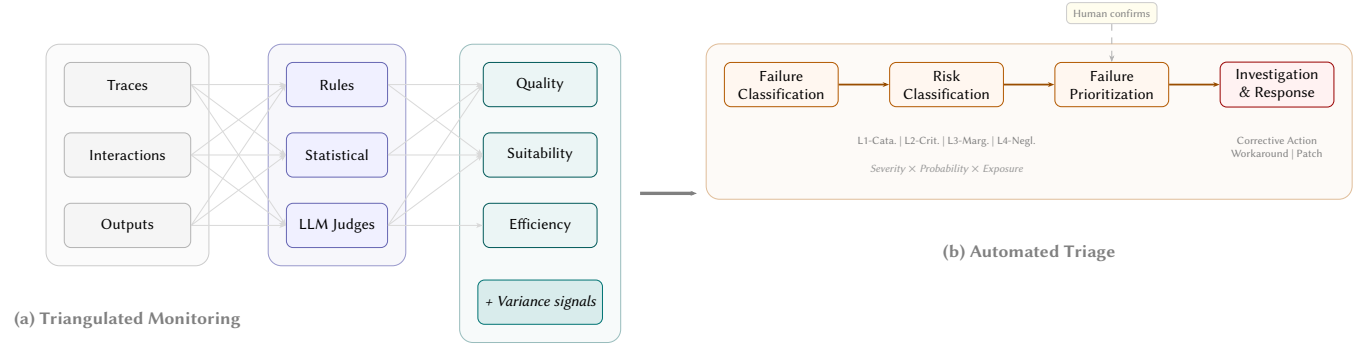

Each dimension is operationalized through a combination of rule-based
validators (e.g., arithmetic checks, schema conformance), statistical
measures (e.g., distributional comparisons across document types),
and LLM-based evaluators that assess semantic properties such as
instruction adherence and output relevance. LLM-based evaluators have
been shown to approximate human preferences with over 80\%
agreement~\cite{zheng2023judging}, though they exhibit systematic
biases including position sensitivity and verbosity preference. This
evaluator variance is one reason within-category variance in our
methodology must be interpreted carefully (see
\S\ref{sec:discussion}). This evaluator diversity provides
robustness: when one evaluator type is unreliable for a given failure
mode, others compensate. Figure~\ref{fig:architecture}(a) illustrates
the monitoring architecture.

\subsection{Variance as a Detection and Characterization Signal}
\label{sec:variance-method}

A key methodological feature of this monitoring methodology is the use
of \textit{within-category variance} as a first-class signal,
complementing aggregate means. Variance serves two purposes depending
on system maturity: in mature systems, it can detect unreliable
behavior that mean performance obscures; in immature systems, it
characterizes the monitoring architecture itself, distinguishing
high-volume routine output from sparse, high-severity findings.

The intuition is straightforward. A system with a high mean score and
low variance is predictable: it performs consistently, and its mean
is a reliable summary of its behavior. A system with a high mean
score and high variance is a liability: its average looks acceptable,
but individual outcomes are unpredictable, and the tails of its
performance distribution may contain operationally significant
failures.

In reliability engineering, this distinction maps directly to the
concept of system maturity~\cite{rausand2004system}. Immature systems
exhibit high variance because failure modes have not yet been
identified and corrected. As corrective actions accumulate, variance
decreases and the system stabilizes. Tracking variance thus provides
a proxy for the system's position on its reliability growth curve,
information that mean performance alone cannot convey.

We operationalize variance through z-scores computed at two
complementary scopes. \textit{Within-run z-scores} use the documents
processed in a single system execution as the reference cohort. For a
metric $x$ measured on document $d$ in a run containing $n$
documents, the z-score is $z_d = (x_d - \bar{x}) / s$, where
$\bar{x}$ and $s$ are the cohort mean and standard deviation.
\textit{Cross-run z-scores} use the fleet of completed runs as the
reference cohort. For a run-level aggregate $\bar{x}_r$, the z-score
is computed via leave-one-out estimation:
$z_r = (\bar{x}_r - \bar{x}_{-r}) / s_{-r}$, where $\bar{x}_{-r}$
and $s_{-r}$ exclude run $r$. This prevents a single anomalous run
from inflating the reference distribution and masking its own
deviation.

A document or run is flagged when its z-score exceeds a threshold
$\tau$ in the direction appropriate to the metric. We set
$\tau = 2.0$ for most evaluators, corresponding to the
${\approx}2.3\%$ tail of a normal distribution. This threshold
balances sensitivity against false positive rate in small cohorts
(typically $n \approx 10$ documents per run). For timing-sensitive
evaluators where right-skewed distributions inflate tail probability,
we use a more stringent $\tau = 3.0$. Flagging direction is
evaluator-specific: quality evaluators flag only downward outliers
(underperformance), while efficiency evaluators flag only upward
outliers (excessive resource consumption). This asymmetry reflects
the operational semantics of each metric---an unusually low
extraction confidence is concerning, but an unusually high one is
not. Every evaluator signal flows directly into the triage system (Section~\ref{sec:triage}), making monitoring active rather than archival.

\section{Automated Triage via FMEA}
\label{sec:triage}

Monitoring generates evidence. But evidence without prioritization
consumes oversight resources indiscriminately: analysts review
low-severity anomalies at the same cost as critical failures, and
the most operationally significant issues compete for attention with
routine noise. This problem is especially acute in agentic systems
where the volume of flagged findings can overwhelm human reviewers
and obscure the failures that demand immediate action.

Triage converts detection into prescription: given a set of
monitored failures, which ones should be addressed first, by whom,
and with what urgency? To answer this systematically, we adapt
Failure Mode and Effects Analysis
(FMEA)~\cite{iec60812, milstd1629a, stamatis1995fmea}, a
methodology developed for safety-critical industries including
aerospace, automotive, and medical devices, to the domain of agentic
AI systems.

\subsection{Four-Phase Triage Process}

Our triage system operates in four phases, each transforming raw
monitoring signals into increasingly actionable outputs.

\textbf{Phase 1: Failure Classification.} Detected failures are
categorized by type and clustered into failure classes.
Classification enables pattern recognition across similar incidents
and distinguishes between isolated anomalies and symptoms of
systemic issues. In our methodology, failure classification draws on
structured metadata from the triangulated monitoring system
(Figure~\ref{fig:architecture}): evaluator scores across quality,
suitability, and efficiency dimensions, rule and policy violations,
error taxonomy labels, tool and runtime errors, and records of human
interventions. This metadata provides substantially richer signal for
classification than single-threshold alerts.

\textbf{Phase 2: Risk Classification.} Classified failures undergo
multi-dimensional risk assessment. Following FMEA methodology, risk
is computed across three axes:

\begin{itemize}
  \item \textit{Severity}: the consequence of the failure on system
  operations, users, and organizational objectives.
  \item \textit{Probability}: the likelihood that the failure will
  occur or recur.
  \item \textit{Scope/Exposure}: the breadth of impact, including the
  range of system components affected and the duration of the
  failure's effects.
\end{itemize}

This multi-factor assessment captures operational and organizational
context that traditional FMEA (which focuses on severity, occurrence,
and detection) does not fully address in AI system settings. The goal
is a systematic, repeatable method that supports objective
prioritization regardless of the specific risk factors chosen.

\textbf{Phases 3 and 4: Failure Prioritization and Response Determination.} Prioritized failures are ranked by urgency and routed to investigation, where root causes are identified and response types (corrective actions, workarounds, or patches) are determined based on technical nature and implementation feasibility.

\subsection{Severity methodology}

Central to the triage process is a severity classification calibrated
to operational impact rather than abstract technical accuracy. We
adapt the four-level methodology from MIL-STD-882E~\cite{milstd882e},
a standard widely used for system safety across defense, aerospace,
and increasingly in software systems:

\begin{itemize}
  \item \textbf{L1 (Catastrophic)}: failure compromises the integrity
  of the overall process; requires immediate escalation and halting of
  automated processing.
  \item \textbf{L2 (Critical)}: failure produces materially incorrect
  results or misses significant findings; requires human review before
  any downstream reliance.
  \item \textbf{L3 (Marginal)}: failure produces degraded but usable
  results; addressed through periodic sampling and quality review.
  \item \textbf{L4 (Negligible)}: failure has minimal operational
  impact; handled automatically or deferred.
\end{itemize}

These categories serve a dual purpose. First, they provide the target
labels for automated triage: given the structured metadata from
monitoring, the system assigns each detected failure to one of the
four severity levels. Second, they define the oversight response: L1
and L2 failures trigger mandatory human escalation, L3 failures
receive periodic sampling, and L4 failures are handled
automatically. This \textit{right-sized} oversight design ensures
that human attention is concentrated where professional judgment and
organizational accountability matter most, rather than distributed
uniformly across all detected anomalies.

This severity-calibrated approach aligns with the NIST AI Risk
Management methodology~\cite{nist2023airmf}, which calls for risk
management practices that are proportionate to the potential impacts
of AI system failures and integrated across the AI lifecycle.

\subsection{From Monitoring to Triage}

The connection between monitoring (\S\ref{sec:methodology}) and triage
is not incidental but architectural. The triangulated monitoring
system produces structured metadata per finding: evaluator scores
across three dimensions, variance signals, error classification
labels, and contextual information about the documents and agent
behavior involved. This metadata serves as the input for severity
classification and governance routing.

Variance signals can provide information that mean-based features
alone cannot. In a mature system, a failure that occurs in a
high-variance cell---where the agentic system sometimes detects the issue and
sometimes does not---carries different risk implications than the
same failure in a low-variance cell where behavior is consistent. In
the current experiment, variance plays a different role: it
characterizes monitor scope classes
(Section~\ref{sec:variance-result}), informing governance design even
before task-level detection is viable.

Figure~\ref{fig:architecture}(b) illustrates the triage
architecture: structured monitoring output feeds failure
classification, risk calculation across the
severity--probability--exposure cube, and prioritized routing to
investigation and response. In \S\ref{sec:results}, we evaluate
whether this architecture produces governance routing that
concentrates human attention on the findings most likely to reflect
genuine system failures.

\section{Experimental Setup}
\label{sec:setup}

\noindent The primary users of this methodology are AI operations teams and the domain practitioners---auditors, compliance leads, clinical reviewers---who bear professional responsibility for outputs produced with agentic assistance and require actionable intelligence about system behavior before they are asked to trust it.  We validate the monitoring and triage methodology on a synthetic
testbed with known ground truth. The testbed is designed to evaluate
whether scope-differentiated monitoring produces actionable
structural diagnosis of an early-stage agentic system, and whether
FMEA-based triage routes the resulting findings to governance
responses calibrated by severity.

\subsection{Synthetic Testbed}

The testbed consists of 120 document bundles generated via
Simthetic for the Subsequent Events and Unrecorded Liabilities. Each bundle represents a complete audit sampling unit for a fictional manufacturing company: a set of 10 interrelated CSV documents including disbursement
listings, expense sub-ledgers, goods and services invoices, purchase
orders, shipping documents (sending and receiving), bank statements,
wire transfers, and purchase contract snippets. Recent work confirms
that synthetic data quality dimensions beyond statistical fidelity
significantly impact agent evaluation
outcomes~\cite{iskander2024quality}.

Simthetic is a synthetic data platform that generates realistic financial audit document bundles with controlled error injection. Synthetic generation is a prerequisite for this experiment's central claim: the masking result (100 injected errors statistically indistinguishable from clean baselines) is only demonstrable with known ground truth. The platform addresses the data access constraint inherent to regulated industries---controlled experiments on client financial documents are not permitted---while producing document structures that reflect the content patterns of real audit workflows.

The testbed is divided into 20 clean baseline bundles containing no
seeded errors and 100 error bundles, each with exactly one injected
error type. The 100 error bundles follow a factorial design: 5 error
subtypes $\times$ 4 difficulty levels (easy, medium, hard,
very\_hard) $\times$ 5 document sets per cell. This design enables
analysis of both cross-subtype and within-subtype variance in agent
behavior. The inclusion of paired clean and error bundles for 100
companies allows direct comparison of monitor firing patterns across
conditions.

Ground truth is documented per bundle in a structured manifest
(\texttt{errors.json}) specifying the error type, affected documents
and fields, and expected detection behavior. All 120 bundles were
validated prior to experimentation using independent ground-truth
checkers: clean bundles contain no seeded anomalies, and every error
bundle's injected error is confirmed present in the expected document
and field.

\subsection{Error Taxonomy}

The five error subtypes span three distinct detection challenges, summarized in Table 1. Arithmetic errors (TotalAmountMismatch, UnitPriceMismatch) require verification of computed values across documents. Cross-document comparison errors (QuantityMismatch, VendorInfoMismatch) require reconciliation of values or identities across independently sourced documents. Temporal anomalies (DateSequenceMismatch) require reasoning about whether document dates are consistent with the audit period. Each subtype is instantiated at four difficulty levels controlling the subtlety of the injected error, from large discrepancies at easy level to near-threshold perturbations at very hard level.

\subsection{Agentic System Under Test}

The purpose of this experiment is not to evaluate a high-performing
agentic system. It is to evaluate whether the monitoring and triage methodology
produces actionable output when applied to the class of systems
where monitoring is most needed: partially integrated, structurally
imperfect, and operating at a maturity level where failure patterns
have not yet been systematically identified.

The agentic system under test has two known integration defects that produce
a characteristic signal pattern. First, document type is identified
at ingestion but not propagated to downstream LLM stages; the field
extractor applies a single hardcoded invoice schema to all document
types regardless of their actual structure. Second, the
cross-document validator receives zero related documents as context,
causing comparison findings to fire without actual cross-referencing.
These are integration gaps, not model failures, and they are typical
of systems deployed before components have been fully connected.

The resulting signal profile is distinctive: the agentic system produces no
actionable error-detection findings. The business resolver reports
``no findings generated'' for every document regardless of whether
errors are present, and the cross-document validator reports
``pass'' uniformly. There is no agent-level detection signal to
score against ground truth. This zero-signal baseline is the
condition the monitoring methodology is designed to operate under.

The system's telemetry nonetheless carries structure. The integration
defects produce consistent processing patterns---the same schema
misapplication, the same manifest gaps---across all runs. But
process-level measurements (extraction completeness, field
confidence, stage timing) vary across documents and runs.
The monitoring methodology operates on this telemetry, not on the
system's own findings.

\subsection{Evaluators and Analysis}

The monitoring methodology deploys evaluators across all three
dimensions described in Section~\ref{sec:methodology}. In this experiment, all evaluators are rule-based or statistical;
no LLM-based evaluators are used, eliminating evaluator variance
as a confound at the cost of departing from typical production
deployments where LLM-based judges are standard. This is a deliberate
methodological choice: introducing evaluator variance would
confound the interpretation of monitor firing variance, which is
the central measurement of this experiment. Production deployments
of the methodology will incorporate LLM-based evaluators within
the Quality and Suitability dimensions; the impact of evaluator
variance on monitor signal interpretation is left for future work. The monitoring layer operates post-hoc on agentic telemetry and does not sit in the critical path of the system's runtime. Computational overhead is a function of telemetry volume and evaluator complexity, both bounded by the scope of the run being analyzed. Statistical evaluators compute within-run z-scores using
the documents processed in a single execution as the reference
cohort, or cross-run z-scores using leave-one-out estimation
against the fleet ($\S$\ref{sec:variance-result}). Flagging direction is
evaluator-specific: quality evaluators flag downward outliers
(underperformance), while efficiency evaluators flag upward
outliers (excessive resource consumption). Every evaluator signal
flows directly into the triage system (Section~3).
 
\begin{table}[t]
\caption{Evaluator suite. Type: R = rule-based, S = statistical.
$\tau$: z-score threshold for flagging. Direction: $\downarrow$ =
flags underperformance, $\updownarrow$ = flags both directions.
Suggested severity is the evaluator's input to the triage system.}
\label{tab:evaluators}
\centering
\small
\begin{tabular}{@{}llllccc@{}}
\toprule
ID & Dim. & Scope & Type & $\tau$ & Dir. & Sev. \\
\midrule
DM-095 & Qual. & Within & R & --- & $\downarrow$ & L3 \\
DM-120 & Qual. & Within & S & 2.0 & $\downarrow$ & L3 \\
DM-160 & Qual. & Within & S & 2.0 & $\downarrow$ & L3 \\
DM-161 & Qual. & Cross  & S & 2.0 & $\downarrow$ & L2 \\
DM-100 & Suit. & Struct.& R & --- & $\downarrow$ & L2 \\
DM-170 & Suit. & Within & S & 2.0 & $\downarrow$ & L3 \\
DM-070 & Eff.  & Within & S & 3.0 & $\updownarrow$ & L3 \\
DM-130 & Eff.  & Within & S & 2.0/3.0 & $\updownarrow$ & L3 \\
\bottomrule
\end{tabular}
\vspace{-1em}
\end{table}

\textbf{Analysis.} The 120 companies produce 220 runs: 100 companies
with paired clean and error runs, and 20 with a clean run only. For
each run, we record the full set of evaluator findings. We then
characterize monitor behavior along three axes:

\begin{enumerate}
  \item \textbf{Firing profiles by scope}: for each monitor, the
  fraction of runs producing $\geq$1 finding, the mean and standard
  deviation of findings per run, and the coefficient of variation
  (CV) across the 220-run fleet.

  \item \textbf{Clean vs.\ error comparison}: for the 100 companies
  with paired runs, whether monitor firing patterns differ between
  clean and error conditions.

  \item \textbf{Triage routing}: the governance response assigned to
  each finding by the deterministic triage rules
  (Section~\ref{sec:triage}), aggregated by scope class and severity
  level.
\end{enumerate}

\section{Results}
\label{sec:results}

We report results from 220 agentic system runs across the 120-company
synthetic testbed described in Section~\ref{sec:setup}: 100 companies
with paired clean and error runs, and 20 with a clean run only. The
monitoring methodology produces 10{,}453 total findings. Three results
emerge: monitor scope determines failure type and severity profile;
variance encodes scope class rather than error presence; and
deterministic triage routes findings to governance responses that
concentrate human attention on approximately 2\% of total output.

\subsection{Monitor Scope Determines Failure Type}
\label{sec:scope}

\begin{figure}[t]
\centering
\includegraphics[width=\columnwidth]{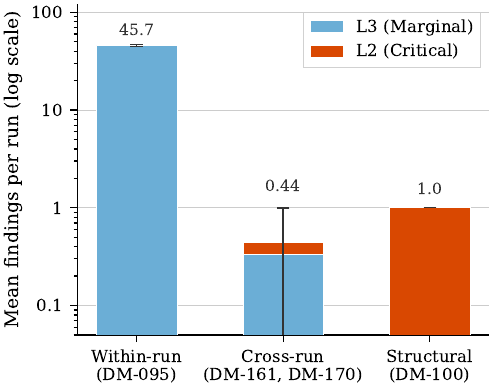}
\caption{Monitor firing patterns by scope across 220 runs.
Within-run monitors (DM-095) produce 45.7 findings per run
($\pm$1.0~sd), all at severity L3 (Marginal). Cross-run monitors
(DM-161, DM-170) fire on 41\% of runs with 24\% of findings at
severity L2 (Critical). Structural monitors (DM-100) fire exactly
once per run, all at L2. Total: 10{,}453 findings across 10{,}136
within-run, 97 cross-run, and 220 structural.}
\label{fig:scope-firing}
\end{figure}

The three scope classes defined in Table~\ref{tab:scope-profiles}
produce sharply different behavioral signatures.
Table~\ref{tab:scope-profiles} summarizes firing profiles and
distributional properties; Figure~\ref{fig:scope-firing} visualizes
the volume and severity contrast.

\textbf{Within-run monitors.}
DM-095 (Extraction Grounding) fires on every run, producing
45.7 findings per run ($\pm$1.0~sd). Output is voluminous, uniform
(CV\,=\,0.02), and entirely L3 (Marginal). This monitor diagnosed a
deterministic stage defect: the field extractor applies a hardcoded
invoice schema to all document types, producing grounding failures
on every non-invoice document regardless of whether errors are
present. The near-zero variance confirms the defect is systematic
rather than stochastic.

\textbf{Cross-run monitors.}
DM-161 (Cross-Run Extraction Completeness) and DM-170 (Stage
Time Concentration) fire on 41\% of runs, producing 0.44 findings
per run across the fleet ($\pm$0.55~sd; CV\,=\,1.25). When they fire,
76\% of findings are L3 and 24\% are L2 (Critical). The high
coefficient of variation reflects the stochastic nature of these
findings: most runs produce zero, a minority produce one or two.
These monitors surface the variable downstream consequences of the
structural integration gaps---runs where the missing context
produces inconsistent extraction completeness or anomalous stage
timing.

\textbf{Structural monitors.}
DM-100 (Document Manifest Gap) fires exactly once per run, on
every run, with zero variance (CV\,=\,0.00). All findings are L2
(Critical).\footnote{DM-100 fires universally because the testbed
simulator includes macOS resource fork files (\texttt{.\_} prefixed)
in the declared document count, inflating the expected count from 10
to 20 while the agentic system processes only the 10 real documents. This is
a simulator artifact, not a system failure, but the monitor correctly
identifies the manifest-to-coverage discrepancy.} This monitor
identified an integration integrity gap: documents expected by the
manifest were not processed by downstream stages. The deterministic,
invariant pattern confirms a structural defect rather than a
stochastic failure.

\textbf{Volume contrast.}
In total, the evaluators produce 10{,}453 findings: 10{,}136
within-run (10{,}057 from DM-095, 61 from DM-120, 18 from DM-160),
97 cross-run (23 from DM-161, 74 from DM-170), and 220 structural
(DM-100). A scope-unaware system would bury the 23 cross-run L2
findings under more than 10{,}000 within-run L3 findings.

\begin{table}[t]
\centering
\small
\caption{Monitor firing profiles by scope (220 runs). CV captures
distributional concentration: within-run output is near-identical
across runs; cross-run output is highly variable.}
\label{tab:scope-profiles}
\setlength{\tabcolsep}{3.5pt}
\begin{tabular}{@{}llrlcl@{}}
\toprule
\textbf{Scope} & \textbf{Eval.} & \textbf{Firing} & \textbf{Find./Run} & \textbf{CV} & \textbf{Sev.} \\
\midrule
Within-run  & DM-095    & 100\% & 45.7\,$\pm$\,1.0$\phantom{^{*}}$  & 0.02 & L3 \\
Cross-run   & -161, -170 &  41\% &  0.44\,$\pm$\,0.55$^{*}$ & 1.25 & L3/L2$^{\dagger}$ \\
Structural  & DM-100    & 100\% &  1.0\,$\pm$\,0.0$\phantom{^{*}}$  & 0.00 & L2 \\
\midrule
\multicolumn{4}{@{}l}{\textit{Total findings}} & \multicolumn{2}{r@{}}{10{,}453} \\
\bottomrule
\end{tabular}

\vspace{3pt}
{\raggedright\footnotesize
$^{*}$All 220 runs; firing runs only: 1.07\,$\pm$\,0.25.\;
$^{\dagger}$76\% L3, 24\% L2.\;
CV\,=\,SD/Mean. DM-120 (61 find./59 runs) and DM-160
(18/9) omitted; both L3.\par}
\end{table}


\subsection{Structural Defects Mask Task-Level Errors}
\label{sec:variance-result}

The experimental design hypothesized that within-category variance
would distinguish error-containing runs from clean baselines. This
hypothesis was not supported. Across the 100 companies with paired
clean and error runs, the two conditions are indistinguishable in
monitor firing patterns. The injected task-level errors are masked
by the system's structural noise floor: when a stage applies the
wrong schema to every document, the resulting findings are identical
whether or not the underlying data contains an anomaly. This is an
instance of evaluation
entanglement~\cite{Boston2026scenariolevel}: when systematic
infrastructure defects dominate, metrics reflect infrastructure
state rather than the task-level signal they were designed to
measure.

\textit{Scope condition.} The masking result observed here depends
on the structural defect producing uniform, content-independent
behavior: the field extractor applies the same incorrect schema to
every document regardless of content, and the cross-document
validator receives zero context on every run. Structural defects
with this property---deterministic and content-independent---are
expected to mask task-level signal completely, as observed.
Structural defects that degrade behavior only conditionally (e.g.,
a context window that truncates only on long documents, or a tool
call that fails only under specific input patterns) would produce
a different pattern: partial masking, in which task-level signal
remains detectable on the subset of inputs where the structural
defect does not fire. Characterizing the full taxonomy of
structural defect types and their masking profiles is left for
future work. What this experiment establishes is the existence and
operational significance of the masking effect for the class of
defects most common in early-stage agentic system
integration~\cite{rausand2004system}.

Variance does, however, encode scope class. The CV column in
Table~\ref{tab:scope-profiles} quantifies the contrast: within-run
monitors produce near-identical output across runs (CV\,=\,0.02),
cross-run monitors produce highly variable output (CV\,=\,1.25), and
the structural monitor produces perfectly invariant output
(CV\,=\,0.00). This distributional signature distinguishes monitor classes without requiring ground truth labels, and is designed for production environments where clean and error conditions are unknown.

\subsection{Triage Routes Findings to Calibrated Governance}
\label{sec:triage-results}

The FMEA-based triage system (Section~\ref{sec:triage}) assigns
governance responses using severity level and monitor metadata.
Table~\ref{tab:triage-routing} presents the routing matrix.

All 10{,}210 L3 findings are routed to \textit{Monitor} (automated
tracking, no human review). All 243 L2 findings---23 from cross-run
monitors and 220 from the structural monitor---are routed to
\textit{Assign Observability} (human investigation). This is a
43$\times$ reduction in the volume requiring analyst attention:
from 10{,}453 total findings to 243.  Notably, 220 of the 243 L2 findings originate from DM-100, which fires on a simulator artifact rather than a system defect. Excluding DM-100, the triage reduction is 10{,}233 / 23 = 445$\times$, carried entirely by the cross-run findings from DM-161. The 43$\times$ figure reported above is the conservative estimate; the methodology's triage value is not dependent on the structural monitor.

The routing is deterministic given severity classification. In this
experiment, routing operates on Severity alone; the experiment does
not evaluate routing under the full Severity-Probability-Exposure
risk cube. The full FMEA methodology (Section~\ref{sec:triage}) defines risk across three axes: Severity, Probability, and Exposure. Probability and Exposure require deployment context that a single-system synthetic testbed does not provide: Probability becomes a meaningful routing input when recurrence patterns can be estimated across system versions, and Exposure becomes differentiating when findings can be attributed to components with bounded scope rather than to defects affecting the entire pipeline uniformly. At Stage~1, Severity directly encodes the operational consequence relevant to governance by prioritizing integration fixes. The methodology incorporates Probability and Exposure as deployment context makes them estimable.

The contribution is the monitoring architecture and severity methodology that make deterministic routing sufficient, eliminating per-incident human classification. The 97\% of findings routed to automated monitoring correspond to the deterministic stage defect identified in Section~\ref{sec:scope}. The 2\% routed to human investigation correspond to the integration integrity gap and the stochastic cross-run consequences that warrant targeted analysis.
\raggedbottom
\begin{table}[t]
\centering
\caption{Triage routing by monitor scope and severity (220 runs).
Deterministic triage routes all L2 findings to human investigation
and all L3 findings to automated monitoring, concentrating analyst
attention on $\sim${\bf 2\%} of total findings.}
\label{tab:triage-routing}
\small
\begin{tabular}{llccc}
\toprule
\textbf{Scope} &
\textbf{Severity} &
\textbf{Triage Action} &
\textbf{Count} &
\textbf{\% of Total} \\
\midrule
Within-run  & L3 & Monitor              & 10,136 & 96.97\% \\
Cross-run   & L3 & Monitor              & 74 & 0.71\% \\
Cross-run   & L2 & Assign Observability & 23 & 0.22\% \\
Structural  & L2 & Assign Observability & 220 & 2.10\% \\
\midrule
\multicolumn{3}{l}{\textit{Total findings}} & 10,453 & 100\% \\
\midrule
\multicolumn{3}{l}{\textit{Routed to Monitor (no human review)}} & 10,210 & 97.68\% \\
\multicolumn{3}{l}{\textit{Routed to Assign Observability (human review)}} & 243 & 2.32\% \\
\bottomrule
\end{tabular}
\parbox{\linewidth}{\footnotesize
Routing is deterministic given severity classification. The final two
rows quantify the governance reduction: the fraction of total findings
requiring human attention.}
\end{table}

\subsection{Summary}

\begin{enumerate}
    \item \textbf{Monitor scope determines failure type and
    severity.} Within-run, cross-run, and structural monitors produce
    categorically different output with different severity profiles.
    Treating them uniformly buries 243 L2 findings under 10{,}210 L3
    findings.

    \item \textbf{Task-level errors are masked by structural defects.}
    Clean and error runs are indistinguishable: task-level errors are
    masked by the structural noise floor. But variance across monitor
    outputs (CV range: 0.00-1.25) characterizes scope class without
    ground truth labels.

    \item \textbf{Deterministic triage concentrates human attention.}
    Scope-aware monitoring and severity classification enable
    rules-based routing that reduces human review volume by
    43$\times$, from 10{,}453 findings to 243.
\end{enumerate}

\section{Discussion}
\label{sec:discussion}

\subsection{What Monitoring Produces at Stage One}

The monitoring methodology produced a correct, actionable diagnosis of
the agentic system's integration state. Within-run monitors
identified a deterministic stage defect: the field extractor
applying a hardcoded invoice schema to all document types regardless
of structure, producing 45.7~findings per run with zero variance
across clean and error conditions. The structural monitor identified
an integration integrity gap: documents dropped between ingestion
and processing, firing exactly once per run on every run. Cross-run
monitors surfaced the stochastic downstream consequences of these
defects, firing on 41\% of runs where the integration gaps produced
inconsistent behavior, with 24\% of those findings classified at
severity L2.

This is a layered structural diagnosis. The three monitor scopes did
not produce interchangeable output. They identified three
structurally distinct failure classes, each with different
operational characteristics: deterministic defects (high-volume,
low-variance, low-severity), integrity gaps (low-volume,
zero-variance, high-severity), and variable consequences
(low-volume, high-variance, mixed severity). The triage system
correctly routed each class to the governance response calibrated
to its operational significance.

The agentic system could not detect injected task-level errors at this
maturity level. This is consistent with what reliability engineering
predicts for immature systems~\cite{rausand2004system}: when
structural failure modes dominate, they must be resolved before
subtler failure detection becomes viable. An agentic system that applies
the wrong extraction schema to every document cannot simultaneously
be expected to detect a 2.5\% arithmetic discrepancy in one of them.
The monitoring methodology correctly prioritized the structural
failures because they are, in fact, the higher-priority problems.

The principle generalizes beyond monitoring. Prior work on synthetic evaluation environments~\cite{Boston2026scenariolevel} demonstrates an analogous entanglement: when evaluation data contains systematic cross-document inconsistencies, agent metrics become uninterpretable because it is impossible to distinguish agent failures from environment defects.

\subsection{Maturity Staging and the Reliability Growth Path}
 
These results instantiate what we propose is the first stage of a
broader reliability improvement
process~\cite{bfg2025relrep}. The maturity-staging model
presented here is validated empirically at Stage~1 only; Stages~2
and 3 are theoretical extrapolations that require longitudinal
study of a system as integration defects are resolved. We present
the staging as a hypothesis the methodology generates, with the
caveat that the transition behaviors described for Stages~2 and 3
are predictions, not observations.
 
\textbf{Stage~1: System characterization} (validated here). The
system is immature; structural defects dominate. Monitoring produces
a diagnostic map: which stages are broken, which connections are
missing, whether each defect is deterministic or stochastic.
Governance focuses on prioritizing integration fixes.
 
\textbf{Stage~2: Error detection.} Integration defects have been
repaired. Monitoring transitions to detecting errors in agent
output. The variance signal that currently encodes scope class
(Table~\ref{tab:scope-profiles}) may at this stage encode behavioral
inconsistency---where the agent sometimes catches an error and
sometimes does not. Concretely: once the hardcoded schema defect
is repaired, DM-095 findings should drop from 45.7 per run to near
zero, and DM-161 becomes the primary signal carrier for task-level
detection failures.
 
\textbf{Stage~3: Reliability tracking.} The system is mature.
Monitoring feeds reliability growth analysis: tracking failure
frequency, Mean Time to Repair, and Fix Effectiveness
Rate~\cite{bfg2025relrep}. The severity classifications
validated here provide the structured input that reliability growth
models require.
 
The staging model implies that monitoring infrastructure should be
deployed early, not after the system matures. The diagnostic output
of Stage~1 identifies which integration points to fix first and
concentrates human attention on variable behavior rather than
deterministic noise. Organizations that defer monitoring until
error detection is needed miss the phase where it delivers the
most immediate engineering value.

\subsection{Generalizability Beyond Audit}
The architecture evaluated here matches a recognizable class of
agentic deployment: document-driven, multi-stage workflows in
regulated industries, including insurance claims, legal contract
analysis, medical records review, procurement verification, and
regulatory compliance. We claim architectural transfer within this
class: the three-scope taxonomy applies wherever systems have
sequential stages, a fleet of runs, and an integration manifest;
the FMEA severity model adapted from
MIL-STD-882E~\cite{milstd882e} applies wherever failures have
operational consequences that can be calibrated to four levels of
oversight response. We do not claim that the specific firing
profiles, CV ranges, or severity distributions reported here
generalize to other systems---these depend on the particular
integration defects present and require domain-specific
calibration. Validating transfer across systems with different
defect profiles, and across non-document-driven agentic
architectures, is the natural next empirical step.

The contribution is more precisely characterized as a design methodology than a framework: no reusable artifact is released. What transfers across domains is the three-scope taxonomy and its mapping to failure type, CV as a scope-characterization signal without ground truth, the FMEA severity classification adapted from MIL-STD-882E, and the triage routing architecture. What must be rebuilt for each new domain are the evaluators themselves, severity calibration to domain-specific business rules, and z-score thresholds calibrated to baseline behavior.

\subsection{Limitations}

Several limitations constrain the current findings. The evaluation uses a synthetic testbed that cannot fully replicate real-world agentic system interactions. The five error subtypes span one audit procedure (SURL under US GAAP); a broader taxonomy would strengthen generalization. Results reflect one agentic system with two specific integration defects; replication across systems with different defect profiles is needed. The maturity-staging model is supported by Stage 1 data only; longitudinal study of a system as defects are repaired is needed to validate transitions between stages.


\section{Conclusion}
\label{sec:conclusion}

We presented a monitoring and triage methodology grounded in
reliability engineering for agentic AI systems in regulated
domains. The methodology decomposes system evaluation into three
dimensions (quality, suitability, efficiency) at three monitoring
scopes (within-run, cross-run, structural), and routes findings
through FMEA-based severity classification calibrated to operational
impact.

Validation on 220~runs of an agentic system processing a
synthetic audit testbed produced a layered structural diagnosis of
the system's integration state. Within-run monitors identified
deterministic stage defects. Structural monitors identified
integration integrity gaps. Cross-run monitors surfaced the
stochastic downstream consequences. Deterministic triage concentrated human investigation on the approximately 2\% of findings reflecting integration integrity gaps and variable downstream behavior, routing the
remaining 97\% to automated tracking.

These results demonstrate that for immature agentic systems,
monitoring provides operational value before error detection becomes
viable. We propose a maturity-staging model in which monitoring
transitions from system characterization to error detection to
reliability tracking as the system matures, and argue that this
progression applies broadly to the document-driven, multi-stage
agentic architectures increasingly common across regulated
industries.

Deploy monitoring early. The first thing it finds is the most
important thing to fix.


\bibliographystyle{ACM-Reference-Format}
\bibliography{reinsai}

\end{document}